\documentclass[12pt]{article}
\usepackage{latexsym}
\usepackage{amsmath}
\usepackage{amsfonts}
\usepackage{amssymb}
\usepackage{amscd}
\usepackage{bbm}
\usepackage{fancybox}
\usepackage{cite}
\usepackage{amsmath,amsfonts,amsbsy}
\usepackage{pstricks,pst-node}
\usepackage[small,bf,hang]{caption2}
\usepackage{graphicx}
\usepackage{epsfig}
\usepackage{psfrag}
\usepackage{comment}

\usepackage{float}

\psset{unit=1.3cm,linewidth=.5pt,radius=.2}  

\usepackage{multirow}                     
\usepackage{float}                          
\usepackage{lscape}                         
\usepackage{bm}

\newcommand{\cD}{{\cal D}}

\newcommand{\cL}{{\cal L}}

\newcommand{\cO}{{\cal O}}

\newcommand{\beq}{\begin{equation}}
\newcommand{\eeq}{\end{equation}}
\newcommand{\bi}{\begin{itemize}}
\newcommand{\ei}{\end{itemize}}
\newcommand{\bt}{\begin{tabular}}
\newcommand{\et}{\end{tabular}}
\newcommand{\bc}{\begin{center}}
\newcommand{\ec}{\end{center}}

\newcommand{\be}{\begin{equation}}
\newcommand{\ee}{\end{equation}}
\newcommand{\bea}{\begin{eqnarray}}
\newcommand{\eea}{\end{eqnarray}}
\newcommand{\ba}{\begin{array}}
\newcommand{\ea}{\end{array}}

\def\bbox{{\,\lower0.9pt\vbox{\hrule \hbox{\vrule height 0.2 cm
\hskip 0.2 cm \vrule height 0.2 cm}\hrule}\,}}
\newcommand{\dsl}{\pa \kern-0.5em /}

\begin{document}

\begin{titlepage}
\begin{center}

\rightline{ UG-14-09}
\rightline{ TUW-14-01}

\vskip 1.5cm

{\Large \bf Logarithmic AdS Waves and Zwei-Dreibein Gravity}

\vskip 1cm

{\bf Eric A.~Bergshoeff$^1$, Andr\'es F. Goya$^{1,2}$, Wout Merbis$^1$ and Jan Rosseel$^3$}

\vskip 25pt

{\em $^1$ \hskip -.1truecm  Centre for Theoretical Physics,
University of Groningen, \\ Nijenborgh 4, 9747 AG Groningen, The
Netherlands \vskip 5pt }

{email: {\tt E.A.Bergshoeff@rug.nl,  w.merbis@rug.nl}} \\
\vskip 10pt

{\em $^2$ \hskip -.1truecm  Universidad de Buenos Aires FCEN-UBA and IFIBA-CONICET, \\
Ciudad Universitaria, Pabell\'on I, 1428, Buenos Aires \vskip 5pt }
{email: {\tt af.goya@df.uba.ar}} \\
\vskip 10pt

{\em $^3$ \hskip -.1truecm  Institute for Theoretical Physics, Vienna University of Technology, \\
Wiedner Hauptstrasse 8–10/136, A-1040 Vienna, Austria\vskip 5pt }
{email: {\tt rosseelj@hep.itp.tuwien.ac.at}} \\
\vskip 10pt

\end{center}

\vskip 1.5cm

\begin{center} {\bf ABSTRACT}\\[3ex]
\end{center}

We show that the parameter space of Zwei-Dreibein Gravity (ZDG) in $\mathrm{AdS}_3$ exhibits critical points, where massive graviton modes coincide with pure gauge modes and new `logarithmic' modes appear, similar to what happens in New Massive Gravity. The existence of critical points is shown both at the linearized level, as well as by finding AdS wave solutions of the full non-linear theory, that behave as logarithmic modes towards the AdS boundary. In order to find these solutions explicitly, we give a reformulation of ZDG in terms of a single Dreibein, that involves an infinite number of derivatives. At the critical points, ZDG can be conjectured to be dual to a logarithmic conformal field theory with zero central charges, characterized by new anomalies whose conjectured values are calculated.

\end{titlepage}

\newpage

\section{Introduction}

Theories of massive gravity in three dimensions have received a lot of attention in recent years. Contrary to ordinary 3D Einstein-Hilbert gravity, they contain propagating bulk gravitons, albeit massive ones. Three-dimensional gravity has often been considered as an interesting toy model for quantum gravity and massive gravity theories open up the perspective of finding toy models for quantum gravity, in the presence of propagating spin-2 particles.

Most of the 3D massive gravity theories considered so far are naturally formulated as higher-derivative theories, i.e. their action consists of an Einstein-Hilbert term, supplemented with a three-derivative Lorentz-Chern-Simons term (introduced in \cite{Deser:1981wh,Deser:1982vy}) and a particular curvature squared term (introduced in \cite{Bergshoeff:2009hq,Bergshoeff:2009aq}). The most general of these models is known as General Massive Gravity (GMG), while particular limits are known as Topologically Massive Gravity (TMG) and New Massive Gravity (NMG). When studied around a Minkowski background, it has been shown that GMG can be made ghost-free, both at the linearized \cite{Bergshoeff:2009hq,Bergshoeff:2009aq} and at the non-linear level \cite{Blagojevic:2010ir,Afshar:2011qw,deRham:2011ca,Hohm:2012vh}. When studied around an AdS background, the presence of the boundary can however introduce subtleties. Indeed, imposing non-trivial boundary conditions for the metric field \cite{Brown:1986nw} can lead to degrees of freedom at the boundary of AdS (often called boundary gravitons), that constitute a two-dimensional conformal field theory. The central charges of the latter are not guaranteed to be positive, if the bulk theory is ghost-free. This is indeed what goes wrong in GMG, that suffers from a tension between requiring absence of ghost-like bulk degrees of freedom and requiring positivity of the central charges of the dual CFT. Recently, it was shown that this tension can be resolved in a different 3D massive gravity model, called `Zwei-Dreibein Gravity' (ZDG) \cite{Bergshoeff:2013xma}. ZDG is a bigravity model, in the spirit of \cite{Hassan:2011zd}, written in first order form as in \cite{Hinterbichler:2012cn}\footnote{See \cite{Chamseddine:1978yu} for earlier related models.}. It describes two Dreibeine, whose dynamics is determined by an Einstein-Hilbert term for each of them, along with particular non-derivative coupling terms. As formulated in this way, it is thus a two-derivative model, that moreover in particular limits can lead to other known 3D massive gravity models, such as NMG and dRGT gravity \cite{deRham:2010kj}. At the linearized level, it propagates two massive helicity-2 modes and it was shown in \cite{Bergshoeff:2013xma,Banados:2013fda} that there exist sectors in the parameter space of ZDG for which the model is ghost-free, both at the linear and non-linear level. Furthermore, it was argued that there exist regions of the parameter space for which bulk ghost-freeness and positivity of the central charges of the dual CFT are compatible. 

Even though unitarity is difficult to achieve, higher-derivative massive gravity theories, such as GMG have found use as potential gravity duals for logarithmic conformal field theories (LCFTs). The latter were introduced in \cite{Gurarie:1993xq} and are typically non-unitary. They are nevertheless often considered in condensed matter physics, e.g. in the context of critical phenomena and percolation. 
As it turns out, at specific `critical points' in the parameter space of these higher-derivative gravity models, some of the linearized modes of these theories degenerate with other modes and are replaced by new so-called `logarithmic modes', that are characterized by different fall-off behavior towards the AdS boundary \cite{Grumiller:2008qz,Liu:2009kc,Liu:2009pha,Alishahiha:2011yb,Bergshoeff:2011ri,Porrati:2011ku}. Theories at such specific parameter values are often denoted as `critical gravities'. Adopting boundary conditions that retain the logarithmic modes and applying the rules of the AdS/CFT correspondence, one finds that the degenerate and corresponding logarithmic modes are dual to operators that constitute a Jordan cell of the Hamiltonian. The CFT Hamiltonian is thus no longer diagonalizable and this feature is the hall mark of an LCFT. Critical gravities can thus be conjectured to give gravity duals of LCFTs and this correspondence has been termed the AdS/LCFT correspondence (see \cite{Grumiller:2013at} for a recent review).

In this paper, we will argue that the existence of critical points in the parameter space, at which logarithmic modes appear in the linearized spectrum, is not only a feature of the higher-derivative gravity models, but also of the recently introduced ZDG model.  We will show this first at the linearized level. We will in particular show that there exist regions in the ZDG parameter space, where the massive gravitons become massless and thus coincide with pure gauge modes. We will show that instead logarithmic modes appear in the linearized spectrum, that behave similarly to the analogous modes found in higher-derivative critical gravities. The existence of these modes can be seen as a serious hint that  ZDG can be added to the class of gravity theories that are, at specific parameter values, dual to LCFTs. We will argue that the dual LCFTs have zero central charges. According to the AdS/LCFT proposal, the degenerate modes and their logarithmic partners are dual to operators, whose two-point functions are governed by quantities, called `new anomalies'. We will calculate the value of these new anomalies on the gravity side via a procedure outlined in \cite{Grumiller:2010tj}.

We will also confirm that the existence of these logarithmic modes is not an artifact of the linearized approximation, but that there exist solutions of the full theory, that exhibit this logarithmic fall-off behavior. This lends further credibility to the idea that the ZDG parameter space features critical points at which the theory can be dual to LCFTs. The non-linear solutions we will discuss are AdS waves, that are the ZDG analogue of the solutions discussed in \cite{AyonBeato:2009yq}. In order to find these solutions, we will use the fact that bimetric theories, such as ZDG, can alternatively be thought of as higher-derivative theories for a single metric field \cite{Hassan:2013pca}. We will express ZDG as a theory whose equations of motion involve an infinite number of derivatives for a single Dreibein and we will show that this allows us to explicitly find AdS waves. We will confirm that AdS waves that exhibit logarithmic asymptotic behavior exist at critical points. 

This paper is organized as follows. In section \ref{sec_linear} we study the ghost-free ZDG model linearized around a maximally symmetric solution. Critical points are discussed and it will be argued that at those points ZDG can be conjectured to be dual to an LCFT with zero central charges. The values of the new anomalies characterizing the dual LCFTs will be calculated. In section \ref{sec_non_linear} we study the non-linear ZDG model and show that full solutions with logarithmic behavior can be found. To do this, we first argue that ZDG can be rewritten as a theory for a single Dreibein involving an infinite number of derivatives. This observation will be used to find AdS wave solutions, that fall off logarithmically at critical points. We end with conclusions and an outlook in section \ref{sec_concl}.

\section{The linear theory}
\label{sec_linear}
In this section we will briefly review the ghost-free ZDG model and consider its linearization. We will discuss the linearized dynamics of fluctuations around a maximally symmetric background and show that there exist critical points in the parameter space where massive modes become massless and logarithmic modes appear, similarly to what happens in critical NMG. The existence of these modes leads one to conjecture that ZDG at such critical points is dual to an LCFT and we discuss the corresponding new anomalies.

\subsection{The ZDG model}
In three dimensions, Zwei-Dreibein gravity can be described as a family of actions with Lagrangian density \cite{Bergshoeff:2013xma}:
\begin{equation}\label{Lbimetric}
\begin{split}
\cL_{ZDG} =  - M_P \bigg\{ & \sigma e_{1\,a} R_1{}^a +  e_{2\,a} R_2{}^a +  \frac{\alpha_1}{6} m^2 \epsilon_{abc} e_1{}^a e_1{}^b   e_1{}^b   
  +  \frac{\alpha_2}{6} m^2 \epsilon_{abc}  e_2{}^a e_2{}^b   e_2{}^c \\ & - \frac12 m^2\beta_1 \epsilon_{abc}  e_1{}^a    e_1{}^b   e_2{}^c - \frac12 m^2\beta_2 \epsilon_{abc}  e_1{}^a    e_2{}^b   e_2{}^c\bigg\} \,.  
\end{split}
\end{equation}
We use a form notation as in \cite{Hohm:2012vh} where products of form fields are to be understood as exterior products.
The basic variables of this model are two Lorentz vector valued one-forms, the Dreibeine $e_{I}{}^a$ with $I= 1,2$, and a pair of Lorentz vector valued connection one-forms $\omega_{I}{}^a$, whose curvature one-forms $R_I{}^a$ are given by:
\begin{equation}\label{Rdefinition}
R_I{}^a = {\rm d} \omega_I{}^a + \frac12 \epsilon^{abc}
\omega_{I\,b} \omega_{I\,c}\,.
\end{equation}
The independent parameters of \eqref{Lbimetric} are two cosmological parameters $\alpha_I$, two coupling constants $\beta_I$ and the Planck mass $M_P$. Besides these, we have introduced a convenient, but redundant parameter $m^2$ with mass-squared dimension and a sign parameter $\sigma = \pm1 $.

A Hamiltonian analysis reveals that, for generic coupling constants $\beta_1$ and $\beta_2$, the ZDG Lagrangian describes 3 bulk degrees of freedom, of which two correspond to the helicity-$\pm 2$ modes of a massive graviton. The third degree of freedom is potentially ghost-like; there do however exist regions of the parameter space where it is removed. An easy way to achieve this, is by restricting the theory to $\beta_1\beta_2 = 0$ and assuming invertibility of one of the Dreibeine \cite{Deffayet:2012nr,Bergshoeff:2013xma,Banados:2013fda}.\footnote{An alternative way to remove the ghost degree of freedom without restricting the parameters of the theory is to assume invertibility of the linear combination $\beta_1 e_1{}^a + \beta_2 e_2{}^a$.} In this work we will choose to work with the $\beta_2=0$ sector of the theory, which is ghost-free for invertible $e_1{}^a$.

The equations of motion for $e_{1}{}^a, e_{2}{}^a$ and $\omega_{I}{}^a$, derived from the Lagrangian density \eqref{Lbimetric}, with $\beta_2 =0$ are given by:
\begin{align}
0  = & \;  \sigma R_1^a + \frac12 m^2 \epsilon^{abc} \left[
 \alpha_1 e_{1\,b}e_{1\,c} - 2 \beta_1 e_{1\,b}e_{2\,c} \right]\,,
\label{R1eom} \\[.3truecm]
0  = &  \;  R_2^a + \frac12 m^2  \epsilon^{abc} \left[
\alpha_2 e_{2\,b}e_{2\,c} - \beta_1  e_{1\,b}e_{1\,c}  \right]\,,
\label{R2eom} \\[.3truecm]
0  = & \; T_{I}^a\,. \label{Teom}
\end{align}
Here, the $T_I^a$ are the torsion 2-forms, given by:
\begin{equation}\label{Tdefinition}
T_I{}^a = \cD_I e_I{}^a
\equiv {\rm d} e_I{}^a + \epsilon^{abc} \omega_{I\,b} e_{I\,c}\,,
\end{equation}
where $\cD_I$ is the covariant derivative with respect to $\omega_I{}^a$ for $I= 1,2$. 
Note that the curvature and torsion 2-forms satisfy the Bianchi identities:
\begin{equation}\label{Bianchi}
\cD_I R_I{}^a = 0\,, \qquad 
\cD_I T_I{}^a = \epsilon^{abc} R_{I\,b} e_{I\,c}\,.
\end{equation}
Each of the kinetic terms of the Dreibeine are invariant under their own diffeomorphisms and local Lorentz transformations. Due to the presence of the interaction term, these symmetries are broken to their diagonal subgroups, defined by identifying the two sets of gauge parameters.

ZDG allows for maximally symmetric vacua, given by:
\begin{align} \label{AdSsol}
e_1{}^a = \bar{e}^a\,, \qquad e_2{}^a = \gamma \bar{e}^a\,, \qquad \omega_I{}^a = \bar{\omega}^a \,,
\end{align}
where $\gamma$ is a scaling parameter. The $\bar{e}^a$ and $\bar{\omega}^a$ are a Dreibein and spin-connection for a maximally symmetric space-time with cosmological constant $\Lambda$ and as such obey:
\begin{align}
& d\bar{\omega}^a + \frac12 \epsilon^{abc} \bar{\omega}_b \bar{\omega}_c - \frac12 \Lambda \epsilon^{abc} \bar{e}_b\bar{e}_c = 0\,,\\
& \bar{\cD} \bar{e}^a \equiv d\bar{e}^a + \epsilon^{abc} \bar{\omega}_b \bar{e}_c= 0\,.
\end{align} 
Indeed, it can then be seen that \eqref{AdSsol} is a solution of the ZDG equations of motion, provided that the scaling parameter $\gamma$ and cosmological constant $\Lambda$ obey:
\begin{align}\label{cc12}
  \alpha_1 =  2 \gamma \beta_1 -\sigma \frac{ \Lambda}{m^2}  \,, && 
 \gamma^2 \alpha_2 =   \beta_1 -  \frac{\Lambda}{m^2} \,. 
\end{align}
For given values of the parameters $\alpha_I$, $\beta_I$, these can be seen as two equations that can be solved to express $\gamma$ and $\Lambda$ in terms of the ZDG parameters.

\subsection{Linearized theory}

We will now linearize ZDG around an AdS$_3$ space-time with cosmological constant $\Lambda$, characterized by the Dreibein $\bar{e}^{a}$ and spin-connection $\bar{\omega}^{a}$ as described above. We thus expand the two Dreibeine and spin-connections, taking the scaling parameter $\gamma$ of \eqref{AdSsol} into account, as follows:
\begin{align}
e_{1}{}^a & =  \bar{e}^a + \kappa h_{1}{}^a\,, &
\omega_{I}{}^a & = \bar{\omega}^{a} + \kappa v_{I}{}^a\,, \\
e_{2}{}^a & = \gamma \left( \bar{e}^a + \kappa h_{2}{}^a \right)\,, \nonumber
\end{align}
where $\kappa$ is a small expansion parameter.  The linear terms in the expansion of the Lagrangian density \eqref{Lbimetric} cancel when eqs. \eqref{cc12} hold.
The quadratic Lagrangian for the fluctuations $h_{I\,\mu}{}^{a}$ and $v_{I\,\mu}{}^a$ is given by:
\begin{align} \label{lin_L}
\cL^{(2)} =  - \sigma M_P & \left[h_{1\,a}\bar{\cD}v_{1}{}^a + \frac12 \epsilon^{abc} \bar{e}_a \left(v_{1\,b}v_{1\,c} - \Lambda h_{1\,b}h_{1\,c}\right) \right] \nonumber \\
 - \gamma M_P & \left[h_{2\,a}\bar{\cD}v_{2}{}^a + \frac12 \epsilon^{abc} \bar{e}_a \left(v_{2\,b}v_{2\,c} - \Lambda h_{2\,b}h_{2\,c}\right) \right] \\
\nonumber  & - \frac12 m^2 \gamma \beta_1M_P \epsilon^{abc}\bar{e}_a \left( h_{1\,b}-  h_{2\,b}\right) \left(  h_{1\,c} -  h_{2\,c}\right) \,.
\end{align}
Provided $\sigma + \gamma \neq 0$, this Lagrangian can be diagonalized by performing the linear field redefinition:
\begin{align} \nonumber
(\sigma + \gamma)  h_{+}{}^a &  = \sigma  h_{1}{}^a +  \gamma  h_{2}{}^a\,, &
h_{-}{}^a  & =  h_{1}{}^a -  h_{2}{}^a\,, \\
(\sigma + \gamma) v_{+}{}^a  & = \sigma v_{1}{}^a +  \gamma  v_{2}{}^a \,, &
v_{-}{}^a & =  v_{1}{}^a -  v_{2}{}^a \,.
  \label{lin_redef}
\end{align}
In terms of these fields the linearized Lagrangian becomes:
\begin{align} \label{linear_bi_lagrangian}
\cL^{(2)}   =  - (\sigma + \gamma) M_P & \left[ h_{+\,a} \bar{\cD} v_{+}{}^a + \frac12 \epsilon_{abc} \bar{e}^a \left( v_{+}{}^b v_{+}{}^c - \Lambda h_{+}{}^{b} h_{+}{}^{c} \right) \right]  \\[.2truecm]
  - \frac{\sigma \gamma }{(\sigma + \gamma)} M_P & \bigg[ h_{-\,a} \bar{\cD} v_{-}{}^a + \frac12 \epsilon_{abc} \bar{e}^a \left( v_{-}{}^b v_{-}{}^c - \Lambda h_{-}{}^{b} h_{-}{}^{c} \right)  + \frac12 M^2 \epsilon_{abc} \bar{e}^a h_-{}^b h_{-}{}^c  \bigg]\,,
\nonumber\end{align}
where the mass parameter $M$ is given in terms of the ZDG parameters as:
\begin{equation}\label{MFP}
M^2 =  m^2 \beta_1\frac{\sigma + \gamma}{\sigma} \,.
\end{equation}
By solving the equations of motion for $v_{\pm}{}^a$  in terms of the perturbations $h_{\pm}{}^a$  and substituting the result back into the Lagrangian, one can see that \eqref{linear_bi_lagrangian} reduces to the sum of a linearised Einstein-Hilbert Lagrangian for $h_+{}_{\mu\nu} = h_{+\,\mu\, a} \bar{e}_{\nu}{}^a$ and a Fierz-Pauli Lagrangian for $h_-{}_{\mu\nu} = h_{-\,\mu\, a} \bar{e}_{\nu}{}^a$, that describes two massive helicity-$\pm2$ modes with mass $M$ in an AdS$_3$ background.

\subsection{Critical Points}
The diagonalization described above fails when $ \sigma + \gamma = 0$ and the Fierz-Pauli mass in eq.~\eqref{MFP} vanishes. This corresponds to a critical point\footnote{In fact, there is a line of critical points. Indeed, for $\sigma + \gamma = 0,$ the parameter relations \eqref{cc12} reduce to $\alpha_1 = -\sigma \left(2 \beta_1 + \Lambda/m^2\right)$ and $\alpha_2 = \beta_1 - \Lambda/m^2$. For a given value of the cosmological constant, there is thus a free parameter $\beta_1$ left. In the following, we will however keep on using the terminology `critical point', often using the plural form to emphasize that there is a continuous family of critical points in ZDG.} in the ZDG parameter space, where logarithmic modes appear, as we will now show. 

In terms of the fields:
\begin{align} \nonumber
h_{-\,\mu}{}^a  & = m^2 \beta_1 \left( h_{1\, \mu}{}^a -  h_{2\, \mu}{}^a \right) \,, & h_{+\,\mu}{}^a &  =  h_{1\, \mu}{}^a +  h_{2\, \mu}{}^a\,, \\ \nonumber
v_{-\,\mu}{}^a  & = m^2 \beta_1  \left( v_{1\, \mu}{}^a - v_{2\, \mu}{}^a \right) \,,
& v_{+\,\mu}{}^a & = v_{1\, \mu}{}^a +  v_{2\, \mu}{}^a \,.
\end{align}
the Lagrangian \eqref{lin_L} becomes:
\begin{align} \label{crit_bi_lagrangian}
\cL^{(2)}   =  \frac{M_P}{2m^2 \beta_1 } \big(  & h_{+\,a} \bar{\cD} v_{-}{}^a +  h_{-\,a} \bar{\cD} v_{+}{}^a  + \epsilon_{abc} \bar{e}^a \left( v_{+}{}^b v_{-}{}^c - \Lambda h_{+}{}^{b} h_{-}{}^{c} \right)
\\ \nonumber & - \epsilon_{abc} \bar{e}^a h_-{}^b h_{-}{}^c \big)\,.
\end{align}
This Lagrangian corresponds to the first order form of the Lagrangian for linearized critical NMG, where the massive modes degenerate with the massless ones and new, logarithmic solutions appear \cite{Liu:2009kc,Bergshoeff:2011ri}. The only difference with the critical NMG case is the appearance of the coupling constant $\beta_1$ as an overall factor.

The equations of motion derived from the Lagrangian density \eqref{crit_bi_lagrangian} are given by:
\begin{align}\label{crit_eom}
\bar{\cD}v_{-}{}^a - \Lambda \epsilon^{abc} \bar{e}_b h_{-\,c} & = 0\,, \nonumber \\
\bar{\cD}v_{+}{}^a - \Lambda \epsilon^{abc} \bar{e}_b h_{+\,c} & = 2 \epsilon^{abc}\bar{e}_b h_{-\,c}\,, \\
\bar{\cD}h_{\pm}{}^a + \epsilon^{abc} \bar{e}_b v_{\pm\,c} & = 0\,. \nonumber
\end{align}
The last of these equations can be used to express $v_{\pm}{}^a$ in terms of $h_{\pm}{}^a$. One finds:
\begin{equation}\label{vpm}
v_{\pm\,\mu}{}^a(h_{\pm}) = -\det(\bar{e})^{-1}\varepsilon^{\nu\rho\sigma} \left( \bar{e}_{\sigma}{}^a \bar{e}_{\mu\,b} - \frac12 \bar{e}_{\mu}{}^a \bar{e}_{\sigma\, b} \right) \bar{\cD}_{\nu} h_{\pm\,\rho}{}^b \,.
\end{equation}
Furthermore, by acting on the equations \eqref{crit_eom} with $\varepsilon^{\rho\mu\nu}\bar{\cD}_{\rho}$ and using the identity $\bar{\cD} \bar{\cD} f^a = \epsilon^{abc} \bar{R}_b f_c = \frac12 \Lambda \epsilon^{abc} \epsilon_{bde} \bar{e}^d \bar{e}^e f_c $ one can derive the constraints:
\begin{equation}
\varepsilon^{\mu\nu\rho}\bar{e}_{\mu}{}^a \bar{e}_{\nu}{}^b h_{- \,\rho\,b} = 0\,, 
\end{equation}
which imply that the field $h_{-\,\mu\nu}= h_{-\,\mu\,a}\bar{e}_{\nu}{}^a$ is symmetric:
\begin{equation}
	 h_{-\,[\mu\nu]}=0\,. \label{lin_symm_constr}
\end{equation}
Plugging \eqref{vpm} into the first two equations of \eqref{crit_eom} and writing them with free space-time indices we obtain:
\begin{align}\label{lineom}
\mathcal{G}_{\mu\nu}(h_-) = 0\,, && \mathcal{G}_{\mu\nu}(h_+) = h_{-\mu\nu} - \bar{g}_{\mu\nu}h_-\,,
\end{align}
where $\mathcal{G}_{\mu\nu}(h)$ is the linearized Einstein tensor which is invariant under linearized diffeomorphisms by construction:
\begin{equation}
\mathcal{G}_{\mu\nu}(h) = - \frac12 \varepsilon_{(\mu}{}^{\alpha\rho} \varepsilon_{\nu)}{}^{\beta\sigma} \bar{\cD}_{\alpha} \bar{\cD}_{\beta} h_{\rho\sigma} - \frac12 \Lambda (h_{(\mu\nu)} - \bar{g}_{\mu\nu} h)\,.
\end{equation}
One can see that these linearized equations of motion are equivalent to the NMG linearized equations of motion at the critical point, see ref.~\cite{Bergshoeff:2011ri}. In particular, the analysis of the solutions carries over without modification. We can thus conclude that at critical points, linearized ZDG exhibits logarithmic modes with exactly the same properties as the analogous modes in critical NMG. 

We have thus found critical points where the linearized Lagrangian \eqref{crit_bi_lagrangian} and equations of motion \eqref{crit_eom} are equivalent to the critical NMG ones. These critical points thus constitute a generalization of the NMG critical point. NMG can be retreived from ZDG by performing a limiting procedure, outlined in appendix \ref{sec_NMGLim}, but this limit requires starting from the part of the ZDG parameter space where the sign parameter $\sigma = -1$. In contrast, in the above discussion, we have not assumed this and it is possible to find regions in the ZDG parameter space where $\sigma + \gamma = 0$ for positive $\sigma$. The critical points found here are thus indeed more general than the NMG one.

\subsection{New anomaly}

In the above section, we have confirmed the existence of critical points in linearized ZDG, where logarithmic modes appear that have the same properties as the logarithmic modes that appear in critical higher-derivative gravity theories, such as critical NMG. In the NMG case, the appearance of these modes led to the conjecture that the field theory dual to critical NMG is an LCFT with zero central charges, once appropriate boundary conditions are taken into account. The NMG logarithmic modes can be seen to be dual to the logarithmic partners of the stress-energy tensor components in the dual LCFT. Even though the central charges are zero, the two-point functions of the stress-energy tensor modes and their logarithmic partners are non-trivial and determined by  new quantities, called the `new anomalies'. A simple way to calculate these new anomalies on the gravity side was given in \cite{Grumiller:2010tj}.

Similar conclusions hold at the ZDG critical points. Again, the central charges \cite{Bergshoeff:2013xma}:
\begin{equation}
c_{L/R} = 12 \pi \ell M_{P}(\sigma + \gamma)\,,
\end{equation} 
where $\Lambda = -1/\ell^2$, vanish at a critical point. The two-point functions of the $c_{L/R}=0$ LCFT should instead be characterized by the new anomalies $b_{L/R}$. These can be calculated via the limiting procedure of ref.~\cite{Grumiller:2010tj}. In order to do this, we need to know the conformal weights $(h, \bar{h})$ of the operators dual to the left and right moving massless and massive modes, that are present in the linearized spectrum of non-critical ZDG. These weights can be obtained via an analysis similar to the one presented in \cite{Li:2008dq} and are given by:
\begin{align}
& {\rm Left:} \;\; (h_L, \bar{h}_L) = (2,0)\,, \qquad  {\rm Right:} \;\; (h_R, \bar{h}_R) = (0,2)\,, \\
& {\rm Massive}\; {\rm Left:} \;\; (h_{M,L},\bar{h}_{M,L}) = \left(\frac{3}{2} + \frac12 \sqrt{1+\ell^2 M^2}, -\frac12 + \frac12 \sqrt{1+\ell^2 M^2} \right)\,, \\
& {\rm Massive}\; {\rm Right:}\;\; (h_{M,R},\bar{h}_{M,R}) = \left( -\frac12 + \frac12 \sqrt{1+\ell^2 M^2}, \frac{3}{2} + \frac12 \sqrt{1+\ell^2 M^2} \right)\,,
\end{align}
where $M^2$ is given by eqn.~\eqref{MFP}.
The left and right moving new anomalies can then be calculated via \cite{Grumiller:2010tj}:
\begin{equation}\label{blimit}
b_{L/R} = \lim\limits_{\sigma + \gamma \to 0} \frac{c_{L/R}}{h_{L/R}-h_{M,L/R}}\,.
\end{equation}
Evaluating this limit explicitly, we find the critical ZDG new anomalies:
\begin{equation}\label{newanom}
b_{L/R} = - \frac{48 \pi \sigma M_P }{\ell m^2 \beta_1}\,.
\end{equation}
Equality of the new anomalies is due to the fact that ZDG is a parity even theory. In the NMG limit of ZDG (see appendix \ref{sec_NMGLim}) the new anomalies reduce to $b_{L/R}^{\rm NMG} = - 12 \sigma' \ell/ G$, where $\sigma'$ is the NMG sign parameter. This result agrees with the known expression obtained in \cite{Grumiller:2009sn} at the NMG critical point defined by $m^2 = - (2 \sigma' \ell^2)^{-1} $. The difference in the ZDG case is that the new anomaly is a function of the coupling constant $\beta_1$, instead of a fixed combination of $\ell/G$. This again makes clear that the ZDG critical points are a generalization of the NMG critical point.

\section{Non-linear theory}
\label{sec_non_linear}
In this section, we will look at ZDG solutions at the non-linear level, in order to confirm that the above found logarithmic modes are not just an artifact of the linearized approximation. To this end we construct  AdS wave solutions of ZDG and show that at a critical point their wave profiles contain terms with a similar behavior as that of the logarithmic modes. In order to obtain these AdS wave solutions, we will first show that ZDG can be understood as a theory with a single Dreibein that however contains an infinite number of higher-derivative terms. We will argue that on an AdS wave solution, the contributions due to all higher-derivative terms can be summed into a closed expression. The resulting equations of motion reduce to a fourth order differential equation for the wave profile, that can be solved explicitly. 

\subsection{ZDG as a higher-derivative theory}

In \cite{Hassan:2013pca}, it was observed that bimetric theories can be thought of as higher-derivative theories. In this section we make this connection explicit in the ZDG case with $\beta_2 =0$. First we observe that we can solve the equations of motion (\ref{R1eom}) to obtain an expression for $e_2{}^{a}$ in terms of $e_1{}^{a}$. Using the property $\varepsilon^{\rho \sigma \tau} R_{1\, \sigma \tau}{}^a =  \det({e_1}) e_1{}^{\sigma\, a} G_1{}^{\rho}{}_{\sigma}$, we obtain the following expression for $e_2{}^{a}$:
\begin{align}
	e_{2\, \mu}{}^{a} &= \dfrac{\alpha_1}{2\beta_1} e_{1\, \mu}{}^{a} + \dfrac{\sigma}{m^2 \beta_1} S_{1\,\mu}{}^a \,,
	\label{e2}
\end{align}
where $ S_{1\,\mu}{}^a \equiv S_{1\, \mu\nu} e_{1}^{\nu\, a}$ and $S_{1\, \mu\nu} = R_{1\,\mu\nu} - \frac14 R_1 g_{1\,\mu\nu}$ is the Schouten tensor of $g_{1\,\mu\nu} \equiv e_{1\,\mu}{}^a e_{1\,\nu}{}^b \eta_{ab}$. In this approach to ZDG, we identify $g_{1\,\mu\nu}$ with the physical metric, as we had to assume $e_1{}^a$ to be invertible for the absence of ghosts and $e_2{}^a$ can be expressed as a function of $e_1{}^a$ and its derivatives\footnote{Since $e_2$ can be expressed in terms of $e_1$ and its derivatives on-shell, it is possible to find the inverse of $e_2$ as an infinite expansion in $1/m^2$.}. The field $e_2$ represents the higher-derivative content of the theory. It is similar to the auxiliary field in the two-derivative formulation of NMG
in the sense that it can be solved for algebraically upon using the equations of motion.

Using the relation \eqref{e2}, we can solve the torsion equation $T_2{}^a =0$ for $\omega_2{}^a(e_1)$ as a power series in $1/ m^2$. By expressing $\omega_2{}^a$ as:
\begin{equation} \label{seriesomega2}
\omega_{2\,\mu}{}^a = \sum_{n=0}^{\infty} \frac{1}{m^{2n}} \Omega^{(2n)}_{\mu}{}^a\,,
\end{equation}
and solving $T_2{}^a = d e_2{}^a + \epsilon^a{}_{bc} \omega_2{}^b e_2{}^c = 0$ order by order in $1/m^2$ we find:
\begin{equation}
\begin{split}
\Omega^{(0)}_\mu{}^a = & \; \omega_{1\,\mu}{}^a\,, \qquad \Omega^{(2)}_{\mu}{}^a = - \frac{2 \sigma}{\alpha_1}  C_{1\,\mu}{}^a \,, \\
\Omega^{(2k)}_{\mu}{}^a = & \; - \frac{2 \sigma}{\alpha_1} \det(e_1)^{-1} \varepsilon^{\nu \rho \sigma} \epsilon_{bcd} \left(e_{1\,\nu}{}^a e_{1\,\mu}{}^b -\frac12 e_{1\,\nu}{}^b e_{1\,\mu}{}^a \right) \Omega^{(2k-2)}_{\rho}{}^c S_{1\,\sigma}{}^d \,, 
\end{split} \label{Omega}
\end{equation}
for $k >1$. Here  $C_{1\,\mu}{}^a \equiv C_{1\,\mu\nu} e_{1}^{\nu\, a} $ and $C_{1\, \mu\nu} = \det(e_1)^{-1} \varepsilon_{\mu}{}^{\alpha \beta} \cD_{\alpha} S_{1\, \beta \nu}$ is the Cotton tensor associated with $g_{1\,\mu\nu}$. This result enables us to write $R_2{}^a$ as a series in $1/m^2$:
\begin{equation}\label{R2exp1}
R_2{}^a = \sum_{n=0}^{\infty} \frac{1}{m^{2n}} R_2^{(2n)\,a}\,,
\end{equation}
where the coefficients in this expansion are given by:
\begin{equation}\label{R2exp2}
\begin{split}
R_2^{(0)\,a} = R_1{}^a\,, \qquad R_2^{(2)\, a} = -\frac{2\sigma}{\alpha_1}  \cD C_1{}^a \,, \\
R_2^{(2k)\,a} = \cD \Omega^{(2k)\,a} + \frac12 \sum_{i=1}^{k-1}\epsilon^a{}_{bc} \Omega^{(2i)\,b} \, \Omega^{(2k-2i)\,c}\,,
\end{split}
\end{equation}
with $C_{1\,\mu}{}^a \equiv e_{1 \nu}{}^a C_1{}^\nu{}_\mu$. The covariant derivative $\cD$ is defined with respect to $\omega_1{}^a$. Replacing these expressions into the equation of motion \eqref{R2eom} for $e_2{}^a$ we obtain a higher order differential equation for $e_1{}^a$ as a power series in $1/m^2$. In appendix \ref{sec_ZDG_HD} we have computed the terms up to order $1/m^4$. 

The equation of motion for $e_1{}^a$ thus obtained contains an infinite number of terms and  features infinitely many derivatives. Note that higher-derivative actions with more than four derivatives can, unlike ZDG, propagate two or more massive gravitons (see \cite{Bergshoeff:2012ev} for an example). Typically however, in cases where this happens, there are terms with more than four derivatives acting on the metric field. This is not the case here as $\Omega^{(2n)}$ contains contractions of the Cotton tensor with $(2n-2)$ Schouten tensors and hence $R_2$ is an infinite series of terms that are products of terms that have at most four derivatives acting on the metric tensor. The resulting equations of motion thus do contain an infinite amount of derivatives, but the maximum amount of derivatives acting on the metric is four. Also note that in the case of an infinite number of derivatives, the initial value problem and ensuing counting of degrees of freedom is subtle (see e.g. \cite{Barnaby:2007ve} for a discussion on these issues). The above higher-derivative formulation is thus not at odds with the fact that ZDG propagates a single massive graviton.

\subsection{AdS waves}
To study critical behaviour in non-linear ZDG we look for propagating waves on an AdS$_3$ background with logarithmic decay, analogously to the situation in NMG, as studied in \cite{AyonBeato:2009yq}. As an Ansatz we consider  a Kerr-Schild deformation of AdS$_3$:
\begin{equation}
    g_{\mu \nu} = \bar{g}_{\mu \nu} - f(u,y) k_{\mu}k_{\nu} \,,
\end{equation}
where $\bar{g}_{\mu \nu}$ is the AdS background and $k^{\mu}$ is a light-like vector\footnote{We take $k^{\mu} \partial_{\mu} = (y/\ell)\partial_{v}$.}.
The function $f(u,y)$ is the wave profile. Using Poincare coordinates this leads to the following expression for the  AdS$_3$ wave space-time  Ansatz:
\begin{equation}
    ds^2 = \dfrac{\ell^2}{y^2} (-f(u,y)du^2-2 du dv + dy^2) \,.
\end{equation}
We will choose the following Dreibeine for this metric:
\begin{equation}
e^{0}  = \dfrac{\ell}{y} \left( \sqrt{f(u,y)} du + \dfrac{1}{\sqrt{f(u,y)}} dv \right) \,, \quad
e^{1}  = \dfrac{\ell}{y} \frac{1}{\sqrt{f(u,y)}} dv \,, \quad  e^{2} = \dfrac{\ell}{y} dy \,.
\label{e1AdSwave}
\end{equation}
In order to find the AdS wave solutions, we will follow the same procedure that was outlined in the previous section, in which ZDG was rewritten as a higher-derivative theory. We recall that the equation of motion in this formulation is given by eq. \eqref{R2eom}, where $e_2{}^a$ is understood to be written in terms of $e_1{}^a$ as in \eqref{e2}. We thus use \eqref{e1AdSwave} as an ansatz for the Dreibein $e_1{}^a$ and find that $e_2{}^a$, as determined by \eqref{e2}, is  given by:
\begin{equation}
	e_{2}{}^{0}  = g(u,y) du +h (u,y) dv \,, \qquad	e_{2}{}^{1} = p(u,y)du + q(u,y) dv \,, \qquad  e_{1}{}^{2} =s(u,y) dy \,, 
\label{e2sol}
\end{equation}
where:
\begin{align}
	g(u,y) &= \dfrac{1}{2m^2 \ell \beta_1 y \sqrt{f(u,y)}} \left( 2m^2 \ell^2 \gamma \beta_1 f(u,y) + \sigma y \left( \dfrac{\partial}{\partial y } - y \dfrac{\partial^2}{\partial y^2 } \right) f(u,y) \right) \,,  \nonumber \\
	h(u,y) &= \dfrac{\gamma \ell}{y \sqrt{f(u,y)}} = q(u,y) \,, \\
	p(u,y) &= \dfrac{\sigma}{2m^2 \ell \beta_1 \sqrt{f(u,y)}} \left( \dfrac{\partial}{\partial y } - y \dfrac{\partial^2}{\partial y^2 } \right) f(u,y) \,, \quad s(u,y) = \dfrac{\gamma \ell}{y} \,. \nonumber
	\label{e2AdSwave}
\end{align}
The parameter $\gamma$ appearing in these functions can be determined from \eqref{cc12}. In order to write down the equation of motion \eqref{R2eom}, we also need to evaluate the series expansion \eqref{seriesomega2} for $\omega_2{}^a$. Explicitly calculating \eqref{Omega} for this solution, we see that all contributions to $\Omega^{(2n)}_{\mu}{}^a$ with $n>0$ have the same form\footnote{It might seem strange that this expression contains only three derivatives for every value of $n$, in view of the fact that $\Omega_\mu^{(2n)a}$ contains more derivatives for larger $n$, in order to balance the mass dimensions of the corresponding $m^{-2n}$ in the series \eqref{seriesomega2}. For this particular ansatz however, the $\ell$-parameter features as an extra dimensionfull parameter that can be used to balance dimensions and this explains why it is possible that all $\Omega^{(2n)}_{\mu}{}^a$ feature the same number of derivatives. }:
\begin{equation} \label{OmAdSwave}
\Omega^{(2n)}_{\mu}{}^a = \left( \frac{\sigma}{\ell^2 \alpha_1} \right)^n \ell y \frac{\partial^3 f(u,y)}{\partial y^3} k_{\mu} k_{\nu} e_1^{\nu\,a}\,.
\end{equation}
We can then sum all orders of $1/m^2$ into a closed expression for $\omega_2{}^a$. We find:
\begin{equation} \label{summedupom}
\omega_{2\,\mu}{}^a = \omega_{1\,\mu}{}^a - \frac{\sigma \ell y}{\sigma - \alpha_1 \ell^2 m^2} \frac{\partial^3 f(u,y)}{\partial y^3} k_{\mu} k_{\nu} e_1^{\nu\,a}\,.
\end{equation} 
Replacing this into the equation of motion (\ref{R2eom}), we see that the latter reduces to the following fourth order differential equation for the wave profile:
\begin{align} \label{wavediffeq}
	\frac{1}{y^2 \sqrt{f(u,y)}} \left[ y^4 \frac{\partial^4 f(u,y)}{\partial y^4} + 2y^3 \frac{\partial^3 f(u,y)}{\partial y^3 } \right. &- \notag \\
	\left( 1 + M^2\ell^2  \right) \left( y^2 \frac{\partial^2 f(u,y)}{\partial y^2} \right.
	&- \left. \left. y \frac{\partial f(u,y)}{\partial y } \right) \right] = 0 \,.
\end{align}
Here $M^2$ is the Fierz-Pauli mass \eqref{MFP} and we have used the parameter relations \eqref{cc12}. 

The equation \eqref{wavediffeq} can be solved by separation of variables and proposing that the solutions behave polynomially in $y$ : $f(u,y) = \tilde{f}(u) y^n$.
The power $n$ is determined as a solution of the indicial equation:
\begin{equation}
	n(n-2)\left( n(n-2) - M^2 \ell^2 \right) \,.
	\label{ChPolynom}
\end{equation}
In general, this equation has four roots $n = \left\lbrace 0, 2, n_+, n_- \right\rbrace$, with $n_{\pm} = 1 \pm \sqrt{1 + M^2 \ell^2}$.

The generic solution for the wave profile is then:
\begin{equation}
	f(u,y) = f_0(u) + f_2(u) \left(\dfrac{y}{\ell}\right)^2 + f_+(u) \left(\dfrac{y}{\ell}\right)^{n_+} + f_-(u) \left(\dfrac{y}{\ell}\right)^{n_-} \,.
	\label{GeneralSol}
\end{equation}
The constant and the quadratic terms can always be removed by local transformations \cite{AyonBeato:2005qq}. The relevant parts are then given by the terms involving $y^{n_{\pm}}$. At special points in the parameter space the roots $n_{\pm}$ become degenerate, as we will discuss in the next subsection.

Since the expressions for $\Omega^{(2n)\,a}$ are all proportional to each other, the AdS wave solution \eqref{GeneralSol} is not only a solution to the full theory, but it will solve the equations of motion at every order of $\frac{1}{m^2}$, provided that the parameters appearing in $n_\pm$ are properly adjusted.

\subsection{Special points}

At the ZDG critical points, one has that $\sigma + \gamma = 0$ and thus $M^2 =0$. At such points $n_+ = 2$ and $n_- = 0$ and the indicial equation \eqref{ChPolynom} thus has two degenerate solutions, instead of four distinct ones. The order of the differential equation \eqref{wavediffeq} at the critical points is still four however, and four distinct, albeit potentially non-polynomial solutions should still exist. Ignoring the constant and quadratic solutions that can be removed by local transformations, one finds the following solutions:
\begin{equation}
	f_c(u,y) = f_L(u) \ln\left(\dfrac{y}{\ell}\right) + f_{2L}(u) \left(\dfrac{y}{\ell}\right)^2 \ln\left(\dfrac{y}{\ell}\right) \,.
	\label{SpecialSol2}
\end{equation}
One thus finds AdS waves with logarithmic decay at the critical point  and this is a clear sign that the existence of logarithmic modes in critical ZDG persists at the non-linear level and is not merely an artifact of the linearized approximation.

There is another class of special points, where a degeneracy in the indicial equation \eqref{ChPolynom} takes place. At these points, $M^2 = -1 / \ell^2$ and $n_{\pm} = 1$. The indicial equation \eqref{ChPolynom} thus only has three roots, one of which is degenerate. The equation \eqref{wavediffeq} is again still of order four and thus one non-polynomial solution should exist. The following solutions (again ignoring the ones that can be removed by local transformations) are found:
\begin{equation}
	f_s(u,y) =\left(\dfrac{y}{\ell}\right) \left(  f_1(u)  + f_{1L}(u) \ln\left(\dfrac{y}{\ell}\right) \right) \,.
	\label{SpecialSol}
\end{equation}
One thus finds one AdS wave with logarithmic decay at these special points. This solution was also found in the Isham, Salam, Strathdee $f-g$ theory \cite{Isham:1971gm} in \cite{Afshar:2009rg}. 

This point already appears as a special point in NMG and it is known that for this special point, NMG has black hole solutions that are not locally isometric to AdS$_3$ \cite{Bergshoeff:2009aq,Oliva:2009ip}. At this point, the linearized Fierz-Pauli action in AdS$_3$ features an extra gauge invariance, with scalar parameter. Linearized NMG, being a sum of a linearized Einstein-Hilbert and Fierz-Pauli action for two different fluctuations, inherits this linearized gauge invariance. The same holds for ZDG as can be seen from eq. \eqref{linear_bi_lagrangian} and the ensuing discussion. At the linearized level, NMG and ZDG at those critical points thus only propagate one degree of freedom. This however is no longer true at the non-linear level and the extra linearized gauge invariance is an accidental one.

NMG at this point has been dubbed `Partially Massless Gravity' (PMG), as the massive mode becomes partially massless \cite{Deser:1983mm,Deser:2001pe}. In \cite{Grumiller:2010tj}, it was argued that solutions with logarithmic decay appear in PMG, of the type given in \eqref{SpecialSol} and that this can be taken as a sign that the dual field theory is an LCFT. Interestingly, we have found above that also PMG, originally found as a special version of NMG, can be generalized to a class of special points in the ZDG parameter space.

\section{Conclusions and outlook}
\label{sec_concl}

In this paper, we have shown that the parameter space of ZDG around AdS$_3$ has critical points, where solutions with logarithmic fall-off behavior appear, both at the linearized and non-linear level. These critical points and logarithmic solutions are similar to the ones that appear in critical NMG. Although NMG can be retrieved from ZDG in a particular limit, the critical points found here do however not simply correspond to the NMG one, but can rather be seen as a generalization of the NMG critical point.  

Note that the existence of the ZDG logarithmic solutions found in this paper is non-trivial. In both NMG and ZDG, criticality is signaled when massive modes become massless and degenerate with pure gauge modes. In both cases, one can argue via continuity that new solutions, that are not massive nor massless modes, should appear at a critical point. In NMG, that is naturally formulated as a four-derivative theory, this argument is based on the fact that the equations of motion remain fourth order at the critical point, and hence there should still be four distinct linearized modes. Since the massive modes coincide with the pure gauge modes, new solutions that are not massive, nor pure gauge should appear in the spectrum. They turn out to be solutions of a particular fourth-order differential equation, featuring a fourth-order differential operator that is the square of a second order one \cite{Bergshoeff:2011ri} and such equations typically feature the logarithmic modes. In ZDG, the equations of motion at critical points are still a system of coupled second-order differential equations for two metric fields and there should again still be four distinct linearized modes. What is non-trivial in ZDG is that the new modes that appear instead of the massive modes at the critical point are logarithmic. Indeed, the logarithmic behavior is typical for solutions of particular differential equations of order higher than two and ZDG is naturally formulated in terms of coupled second-order equations. In this paper, we have however seen that the new solutions at the ZDG critical points are logarithmic. At the linearized level, this stems from the fact that the linearized critical ZDG equations of motion are the same as the linearized NMG ones. At the non-linear level, we have shown that ZDG can alternatively be rewritten as a higher-derivative theory (involving an infinite number of higher-derivatives) for a single Dreibein and it is this higher-derivative character of ZDG that is ultimately responsible for the existence of AdS waves with logarithmic fall-off behavior at the critical points.

As in the critical higher-derivative massive gravity cases, the existence of logarithmic modes  can be seen as a hint that critical ZDG theories are dual to logarithmic conformal field theories, once appropriate boundary conditions are imposed. In order to show this in more detail, more checks need to be performed however. In particular, precise calculations of two- and three-point functions, as was done for TMG and NMG in \cite{Skenderis:2009nt,Grumiller:2009mw,Grumiller:2009sn}, via e.g. holographic renormalization should be performed. The conjecture can also be shown by calculations of the classical and one-loop partition functions on the gravity side (see \cite{Gaberdiel:2010xv,Bertin:2011jk} for examples in higher-derivative theories) and checking that the results conform with the structure expected for an LCFT. Performing these checks will require an extension of the AdS/CFT holographic dictionary and methods to bigravity theories, such as ZDG and we will leave this for the future.

\section*{Acknowledgements}
It is a pleasure to thank Hamid Afshar, Daniel Grumiller, Olaf Hohm and Paul Townsend for discussions and comments.
The work of J.R. was supported by the START project Y 435-N16 of the Austrian Science Fund (FWF). W.M. is supported by the Dutch stichting voor Fundamenteel Onderzoek der Materie (FOM). A.F.G. is supported by the EuroTANGO II Erasmus Mundus Programme and by the Consejo Nacional de Investigaciones Cient\'ificas y T\'ecnicas (CONICET).

\appendix
\section{NMG Limit} \label{sec_NMGLim}
In our convention for the parameters of the Lagrangian of ZDG \eqref{Lbimetric} the limit to NMG is slightly different from the one given in \cite{Bergshoeff:2013xma}. To flow to NMG, we should take $\sigma = -1$ and make the field redefinition
\begin{equation}\label{NMGlim}
e_{2}{}^a = \gamma e_{1}{}^a + \frac{\lambda}{m^2} f^a\,, \qquad \omega_{2}{}^a = \omega_{1}{}^a - \lambda h^a \,.
\end{equation}
NMG can then be obtained from the flow:
\begin{align}\label{NMGlim2}
M_P(\lambda) & =  \frac{1}{\lambda} M' \,, &&  \gamma (\lambda) = 1 + \sigma' \lambda \,,  \nonumber \\
\alpha_1(\lambda) & = \left( 6 - \frac{\Lambda_0}{m^2}\right) \lambda + \frac{2(1+\sigma'\lambda)}{\lambda} \,,
&& \beta_1(\lambda) = \frac{1}{\lambda} + \lambda\,, \\
 \alpha_2 (\lambda) & = \frac{1}{\lambda} - 2 \sigma' \,, \nonumber
\end{align}
and sending $\lambda \to 0$. Here $\sigma' = \pm 1$ is a new sign parameter and $\Lambda_0$ is a new cosmological parameter. The Lagrangian 3-form \eqref{Lbimetric} then becomes:
\begin{equation}\label{CSNMG}
\begin{split}
\cL_{\rm NMG} = M'  \bigg\{ & - \sigma' e_{1\, a} R_{1}^a + \frac{\Lambda_0}{6} \epsilon^{abc}e_{1\,a}e_{1\,b}e_{1\,c}+ h_a T^a_1  \\
& - \frac{1}{m^2}\left( f_a R_{1}^a + \frac12 \epsilon_{abc}e_1^a f^b f^c \right)   \bigg\}\,.
\end{split}
\end{equation}
This action is the Chern-Simons-like formulation of New Massive Gravity as considered in \cite{Hohm:2012vh}, with a Planck mass $M' = (8 \pi G)^{-1}$.

\section{ZDG as a higher-derivative theory}
\label{sec_ZDG_HD}
In this appendix we give an expression of the ZDG equations of motion as a function of a single metric with higher-derivative contributions up to order $1/m^4$. This can be obtained by substituting \eqref{R2exp1} with coefficients \eqref{R2exp2} into the equation of motion \eqref{R2eom}. The equation of motion is rewritten in a second-order form, using the metric $g_{\mu\nu} = e_{1\,\mu}{}^a e_{1\,\nu}{}^b \eta_{ab}$ to raise and lower indices. The result is:
\begin{equation}
\begin{split} \label{HDeom}
0 = \sqrt{-g} M_P & \left\{ \left(1+ \frac{\alpha_1\alpha_2\sigma}{2\beta_1^2}\right) G_{\mu\nu} - \left( \frac{\alpha_1^2\alpha_2}{4\beta_1^2}-\beta_1\right) m^2 g_{\mu\nu} + \frac{1}{m^2} E_{\mu\nu} + \frac{1}{m^4} F_{\mu\nu} \right. \\
&\quad  \left. + \cO\left(\frac{1}{m^6} \right)  \right\}\,,
\end{split}
\end{equation}
where here and in the following we have omitted the label 1, used to denote the Dreibein.
The symmetric tensors $E_{\mu\nu}$ and $F_{\mu\nu}$ carry terms with four and six derivatives respectively. They are:
\begin{align} 
E_{\mu\nu} =  & - \frac{2\sigma}{\alpha_1} \bigg[ \Box R_{\mu\nu} - \frac14 (g_{\mu\nu} \Box R + \nabla_{\mu} \nabla_{\nu} R ) - 3 R_{\mu\rho}R^{\rho}{}_{\nu} + g_{\mu\nu} R_{\rho\sigma}R^{\rho\sigma} - \frac12 g_{\mu\nu} R^2 \nonumber \\ & + \dfrac{3}{2} R R_{\mu\nu}\bigg]
+\frac{\alpha_2}{2 \beta_1^2} \left[ g_{\mu\nu} R_{\rho\sigma}R^{\rho\sigma} - \frac58 g_{\mu\nu} R^2 + \frac32 RR_{\mu\nu} - 2 R_{\mu\rho}R^{\rho}{}_{\nu} \right] \,, \label{Eterms} \\
F_{\mu\nu} = & \; \frac{4}{\alpha_1^2} \left\{ \nabla^{\rho} \left[ S_{(\mu}{}^{\sigma} \nabla_{\nu)} S_{\rho \sigma} -  S_{(\mu|}{}^{\sigma}\nabla_{\rho} S_{|\nu) \sigma} - S_{\rho}{}^{\sigma} \nabla_{(\mu}S_{\nu) \sigma} + 2 S^{\sigma}{}_{\mu} \nabla_{[\rho} S_{\sigma] \nu}  \right. \right. \nonumber \\
& \qquad \qquad  \left. \left. + S^{\sigma}{}_{\rho} \nabla_{\sigma} S_{\mu\nu} \right] + \nabla^{\rho} S^{\lambda \sigma} \nabla_{[\lambda} S_{\rho] \sigma} g_{\mu\nu} - 2 \nabla^{\rho} S^{\sigma}{}_{\nu} \nabla_{[\sigma} S_{\rho] \mu} \right\} \label{Fterms}.
\end{align} 
We would like to stress that at order $1/m^2$ the above equations of motion cannot be integrated to an action, unless the following relation between the ZDG parameters holds:
\begin{equation} \label{actionrelation}
	-\frac{\sigma}{\alpha_1} = \dfrac{\alpha_2}{2\beta_1^2} .
\end{equation}
If the ZDG parameters are restricted in this way, the $1/m^2$ contributions to \eqref{HDeom}, given explicity in \eqref{Eterms}, can be integrated to an action proportional to $R_{\mu\nu}R^{\mu\nu} - \frac38 R^2$. This combination of $R^2$ terms corresponds to the higher-curvature part of the NMG action. This is not a coincidence as can be seen by explicitly performing the NMG limit. Indeed, after substituting \eqref{NMGlim2} into the coefficients in eqn.~\eqref{HDeom}, we see that the terms at order $1/m^4$ scale as $\lambda$ and hence vanish in the $\lambda \to 0$ limit, while the remaining coefficients become:
\begin{align}
M_P \left(1 + \frac{\alpha_1\alpha_2 \sigma}{2\beta_1^2} \right) = \sigma' + \cO(\lambda)\,, \nonumber \\
- M_P\left( \frac{\alpha_1^2\alpha_2}{4\beta_1^2}-\beta_1\right) m^2 = \Lambda_0 + \cO(\lambda)\,, \\
- M_P\frac{2 \sigma}{\alpha_1} = 1 + \cO(\lambda)\,, \qquad M_P \frac{\alpha_2}{2\beta_1^2} = \frac12 + \cO(\lambda)\,. \nonumber
\end{align}
In particular, the last two equations show that the $\lambda \rightarrow 0$ limit enforces the parameter relation \eqref{actionrelation} 
and as a consequence the NMG equations of motion  \cite{Bergshoeff:2009hq,Bergshoeff:2009aq} that result from eqn.~\eqref{HDeom} in the $\lambda \rightarrow 0$ limit
\begin{equation}
\begin{split}
0 = \sigma' G_{\mu\nu} + \Lambda_0 g_{\mu\nu} + \frac{1}{m^2} & \left[ \Box R_{\mu\nu} - \frac14 (g_{\mu\nu} \Box R + \nabla_{\mu}\nabla_{\nu}R ) - 4 R_{\mu\rho}R^{\rho}{}_{\nu} \right. \\
& \left. + \frac94 RR_{\mu\nu} + \frac32 g_{\mu\nu} R_{\rho\sigma}R^{\rho\sigma} - \frac{13}{16} g_{\mu\nu}R^2 \right] \,,
\end{split}
\end{equation}
can be integrated to an action, even if for the generic ZDG equations of motion \eqref{HDeom} this is not possible order by order in $m^2$. 

Note that the utility of the higher-derivative formulation of ZDG will very much depend on the specific application one has in mind. In this paper, we have used this formulation in order to obtain AdS wave solutions. For other applications, e.g. in AdS/CFT, an action for which the variational principle is well-defined, is required and one will have to resort to the two-derivative Zwei-Dreibein formulation. Note that even if the higher-derivative terms could be integrated to an action, a formulation without higher derivatives is still more useful in order to set up a well-defined variational principle, as is discussed in the NMG case in \cite{Hohm:2010jc}.


\providecommand{\href}[2]{#2}\begingroup\raggedright\endgroup

\end{document}